\documentclass[preprint,aps,showpacs]{revtex4}
\begin{document}
\preprint{UCI-TR 2002-15}
\title{An explicit time variable for cosmology and the matter-vacuum
  energy coincidence.}
\author{Myron Bander\footnote{Electronic address: mbander@uci.edu}
}
\affiliation{
Department of Physics and Astronomy, University of California, Irvine,
California 92697-4575}

\date{May\ \ \ 2002}

\begin{abstract}
  By allowing for non zero vacuum expectation values for some of the fields
  that appear in the Hamiltonian constraint of canonical general relativity
  a time variable, with usual properties, can be identified; the constraint
  plays the role of the ordinary Hamiltonian. The energy eigenvalues
  contribute to the variation of the scale parameter similarly to the way
  matter density does. For a universe described by a superposition of
  eigenstates or by a thermodynamic ensemble the dominant contribution
  comes from energy, or equivalently effective matter density, of the same
  order as the vacuum energy (cosmological constant). This may explain the
  observed ``coincidence'' of these two values.
\end{abstract}
\maketitle

Canonical Hamiltonian gravity, both classical and quantum, is a theory of
constraints. A normal unitary evolution can be obtained only after some
dynamical variable, whose canonical conjugate will play the role of time,
is identified and solved for as a function of the other dynamical
variables. Although this is the path to be followed in this work, it should
be mentioned that it is not the universal approach to the problem of time
in cosmology. In the `no boundary' approach of Hartle and Hawking \cite{HH}
and in the `nucleation from a point' view of Vilenkin \cite{Vilenkin} the
existence or nonexistence of time is ignored and only correlations between
dynamical variables are looked for. In the extrinsic time approach, time is
related to a particular foliation of a three geometry in four dimensional
space-time; some recent works \cite{GS} have used this method. Extensive
reviews may be found in \cite{Kuchar,Isham} and more recently in
\cite{Barvinsky}. As mentioned, in this work we shall look for a dynamical
variable that can be identified as time. Some of the difficulties that this
approach has had are overcome in that we allow for some fields to acquire
non-zero expectation values before the constraint equations are solved.
This will result in a Schr\"odinger equation with the Hamiltonian
constraint playing the role of the Hamiltonian itself. Applying this to the
evolution of the scale $R(t)$ in cosmological models, we find that the
energy eigenvalues $E$ contribute in the same functional form as the matter
density, which varies as $1/R^3$ but is multiplied by $R^3$. A universe,
whose wave function is a superposition of such eigenstates, may result in
the effective matter density and the vacuum energy (cosmological constant)
tracking each other. This would explain the present ``coincidence'' of
these two values \cite{Coinc}.  In obtaining these results we use the
minisuperspace approximation; at each step, however, we show how the terms
we introduce arise from a Lagrangian fully compatible with general
relativity.

Hamiltonian gravity and cosmology are usually treated in the, afore
mentioned, minisuperspace approximation, where the continuous set of
gravitational and other degrees of freedom are reduced to a small number of
collective, spatially constant fields. In the simplest case, the dynamics
of a Robertson-Walker-Friedman universe, with a metric
\begin{equation}\label{RW}
ds^2=N^2(t)dt^2-R^2(t)\left [\frac{dr^2}{1-kr^2}+r^2d\Omega^2\right]\, ,
\end{equation}
are specified by the action for the {\em one }variable $R(t)$
\begin{equation}\label{RWFaction}
S=\int\left[ P_RdR-Ndt\, {\cal H}_G(P_R,R)\right]\, .
\end{equation}
with $P_R$ being the momentum conjugate to $R$ and 
\begin{equation}\label{RWFhamiltonian}
{\cal H}_G=-\frac{G}{3\pi R}P_R^2-\frac{3\pi}{4G}\left(
  kR-\frac{\Lambda R^3}{3}\right)+2\pi^2\rho(R)R^3\, ;
\end{equation}
$G$ is Newton's constant, $\Lambda$ is the cosmological constant and
$\rho(R)$ accounts for the matter and radiation density.
Varying this action with respect to the lapse function $N$ yields the
constraint ${\cal H}_G(P_R,R)=0$ and setting the variation of ${\cal
  H}_G(P_R,R)$ with respect to $P_R$ and $R$ to zero insures the
conservation of the energy-momentum stress tensor; there is no room for
both time and a dynamical variable that depends on it.

To allow for the introduction of time more fields have to appear.
A Lagrangian for a scalar field $\tau$, 
\begin{equation}
{\cal L}_\tau=\int d^4x\sqrt{-g}\, \frac{g^{\mu\nu}}{2}
   \partial_\mu\tau\partial_\nu\tau\, ,
\end{equation}
can be  added to the one for gravity; in the minisuperspace approximation
this changes eq. (\ref{RWFaction}) to
\begin{equation}\label{RWFaction2}
S=\int \left[P_RdR+\pi_\tau d\tau-Ndt\left({\cal
    H}_G+\frac{\pi_\tau^2}{2R^3}\right)\right]\, , 
\end{equation}
where $\pi_\tau$ is the momentum conjugate to $\tau$. One then tries to
make $\tau$ play the role of time for $R$ or vice versa. A serious
drawbacks of this approach is that the resulting Schr\"odinger equation is
hyperbolic and problems similar to those that occur in the first quantized
treatment of the Klein-Gordon equation also occur here.  These problems
would go away if $\pi_\tau$ were to appear {\em linearly}\/ in the
coefficient of $N$ in eq. (\ref{RWFaction2}). This is a goal of this work.

To achieve this end we add to the following to the Lagrangian for pure
gravity:
\begin{eqnarray}\label{addL}
{\cal L}_{t}&=&\int d^4x\left\{\sqrt{-g}\left[  \frac{g^{\mu\nu}}{2} 
  \partial_\mu\tau\partial_\nu\tau + bg^{\mu\nu}\partial_\mu\tau
   \partial_\nu\chi-c(g^{\nu\nu'}g^{\lambda\lambda'}
     g^{\sigma\sigma'}H_{\nu\lambda\sigma}H_{\nu'\lambda'\sigma'}+h^2)^2
     \right ]\right .\nonumber\\
   &+& \left. d\epsilon^{\mu\nu\lambda\sigma}\partial_\mu\chi
     H_{\nu\lambda\sigma}\right\}\, .
\end{eqnarray}
$\tau$ is the field whose spatially constant part will ultimately play the
role of time. $\chi$ is a cyclic field, $0\le\chi<2\pi$ and
$H_{\nu\lambda\sigma}$ is an antisymmetric three indexed tensor field. With
$h^2$ positive the third term in eq. (\ref{addL}) ensures that the spatial
parts of $H$ acquire an expectation value. It should be noted that due to
the $\epsilon^{\mu\nu\lambda\sigma}$ in the coefficient of $d$, no
$\sqrt{-g}$ is necessary for this term and, by design, no
$(\partial\chi)^2$ appears in the Lagrangian; the cyclic nature of the
field $\chi$ can result from $\chi$ being the phase of a regular complex
field whose magnitude is fixed to some vacuum expectation value. A kinetic
energy for $H_{\nu\lambda\sigma}$ can occur as can a potential involving
$\tau$. We will show that this produces a $\pi_\tau$ appearing linearly in
the constraint. 

The minisuperspace approximation to (\ref{addL}) is
\begin{equation}
{\cal L}_{tMS}=\int dt \left[
     \frac{R^3}{2N}\left(\frac{d\tau}{dt}\right)^2+
       b\frac{R^3}{N}\frac{d\tau}{dt}\frac{d\chi}{dt}-cNR^3\left(
        H_{ijk}H_{ijk}/R^6-h^2\right)^2
          +d\frac{d\chi}{dt}\epsilon^{ijk}H_{ijk}\right]\, .
\end{equation}
Replacing $H_{ijk}$ by its vacuum expectation value
\begin{equation}\label{<H>}
<H_{ijk}>=\frac{R^3h\epsilon_{ijk}}{6}\, .
\end{equation}
the Hamiltonian corresponding to ${\cal L}_{tMS}$ is
\begin{equation}\label{H-tMS}
{\cal H}_{tMS}=\frac{N}{b^2R^3}\left[b\pi_\tau(\pi_\chi-dhR^3)-
  \frac{1}{2}(\pi_\chi-dhR^3)^2\right ]\, .
\end{equation}
In the fixed $\pi_\chi=0$ (a small value for the parameter $b$ would favor
$\pi_\chi=0$) sector we recover a term linear in $\pi_\tau$. Rescaling
$\tau$ we obtain, instead of the Wheeler-de-Witt equation, a Schr\"odinger
equation with ${\cal H}_G$ acting as a normal Hamiltonian \cite{addcosm},
\begin{equation}\label{Scheq}
-i\frac{\partial}{\partial\tau}+{\cal H}_G=0\, .
\end{equation}

The contribution of an eigenvalue $E$ of eq. (\ref{Scheq}) for ${\cal H}_G$
corresponding to eq. (\ref{RWFhamiltonian}) cannot be distinguished from
the contributions of matter, $\rho_M(R)\sim 1/R^3$, to the evolution of the
scale parameter $R$; thus we may consider $(E-2\pi^2\rho_M R^3)$ to be the
effective energy eigenvalue, or equivalently $(\rho_M - E/2\pi R^3)$ the
effective matter density. Interesting results obtain for a universe that,
rather then being a state with a definite energy, is a superposition of
such eigenstates
\begin{equation}\label{superpos}
\Psi(R,\tau)=\int dE f(E)\psi_E(R)e^{-iE\tau}\, ;
\end{equation}
$\psi_E(R)$ is an eigenfunction of ${\cal H}_G$ and $f(E)$ describes the
energy wave packet. The possibility now arises that at different $R$'s
different $E$'s dominate the integral in eq. (\ref{superpos}). Should the
$E$ and $R$ dependence $\ln\psi_E(R)$ appear in the combination 
\begin{equation}\label{scaling}
\ln\psi_E(R)=R^3g\left(\frac{\Lambda R^3}{E}\right)\, ,
\end{equation} 
with $g$ some function and with  $\ln[f(E)]$ varying no faster than $c|E|$,
the dominant contribution to $\Psi(R,\tau)$ in eq. (\ref{superpos}) comes
from $E\sim \Lambda R^3$. As was remarked earlier, $E$ contributes to the
evolution of $R$ the same way as $\rho_M R^3$ and thus {\em for any given $R$
  the effective $\rho_M\sim\Lambda$}. Thus the observation that today
$\rho_M\sim\Lambda$ \cite{Coinc} may not be a coincidence but may hold true
at all scale factors $R$ \cite{Griest}.

An example of $\psi_E(R)$ scaling according to eq. (\ref{scaling}) is the
WKB approximation for the observationally favored case of a flat, $k=0$,
universe \cite{Coinc} in an epoch where the radiation density may be
neglected; $\psi_E$ is a combination of
\begin{equation}\label{WKB}
\psi_E^{(\pm)}(R)=\exp\left(\pm \frac{3\pi}{2G}i\int_0^Rdr \sqrt{
   \frac{\Lambda r^4}{3}-\frac{4GEr}{3\pi}}\right )\, .
\end{equation}
The integrals in the exponents of the above have the desired scaling
property. For a universe described by a thermodynamic density matrix at a
temperature $1/\beta$ the dominant contributions to the diagonal density
matrix elements, those that account for the distribution of the scale
factor $R$,
\begin{equation}
\rho(R,R)=\int dE \psi_E(R)e^{-\beta E}\psi^*_E(R)\, ,
\end{equation}
likewise come from $E\sim\Lambda R^3$.


\begin{thebibliography}{widest_entry} 
\bibitem{HH}
J.B. Hartle and S.W. Hawking, Phys.\ Rev.\ {\bf D28} 2960 (1983).
\bibitem{Vilenkin}
A.D. Linde, JETP {\bf 60}, 211 (1984); A. Vilenkin, Phys.\ Rev.\ {\bf D37}
888 (1987). 
\bibitem{GS}
G. Giribet and C. Simeone, gr-qc/0106027; C. Simeone, gr-qc/0108081. 
\bibitem{Kuchar}
K.V. Kucha\v{r}, in {\it Proceedings of the 4th Canadian Conference on
  General Relativity and Relativistic Astrophysics}\/, edited by
G. Kunstatter, D. Vincent and J. Williams (World Scientific, Singapore,
1992).
\bibitem{Isham}
C.J. Isham, in {\it Integrable Systems, Quantum Groups and Quantum Field
  Theories}\/, edited by L.I. Ibort and M.A. Rodriguez (Kluwer Academic
Publishers, London, 1993). [gr-qc/9210011]
\bibitem{Barvinsky}
  A.O. Barvinsky, in {\it Proceedings of the IXth Marcel Grossmann Meeting
    On Recent Developments In Theoretical And Experimental General
    Relativity, Gravitation And Relativistic Field Theories (MG 9) 2-9 July
    2000, Rome, Italy} [gr-qc/0101046].
\bibitem{Coinc}
S. Perlmutter {\it et al}\/. Nature {\bf 391}, 51 (1998);
Astrophys. J. {\bf 517} 565 (1998); A. Riess {\it et al}\/. Astron. J. {\bf
  116}, 1009 (1998); P.M. Garnovich {\it et al}\/. Astophys. J. {\bf 509}
74 (1998). 
\bibitem{addcosm}
The last term in eq. (\ref{H-tMS}) contributes to the cosmological constant. 
\bibitem{Griest}
An alternate suggestion that it is the effective cosmological constant that
varies has recently been made by K. Griest [astro-ph/0202052].

\end{thebibliography}
\end{document}